# Understanding Illicit Drug Use Behaviors by Mining Social Media

Yiheng Zhou, Numair Sani, Chia-Kuei Lee[+] and Jiebo Luo

University of Rochester, Rochester, NY, United States
[+]University of Rochester Medical Center, Rochester, NY, United States

`{yzhou49,nsani}@u.rochester.edu, jluo@cs.rochester.edu`

**Abstract.** Drug use by people is on the rise and is of great interest to public health agencies and law enforcement agencies. As found by the National Survey on Drug Use and Health, 20 million Americans aged 12 years or older consumed illicit drugs in the past few 30 days. Given their ubiquity in everyday life, drug abuse related studies have received much and constant attention. However, most of the existing studies rely on surveys. Surveys present a fair number of problems because of their nature. Surveys on sensitive topics such as illicit drug use may not be answered truthfully by the people taking them. Selecting a representative sample to survey is another major challenge. In this paper, we explore the possibility of using big data from social media in order to understand illicit drug use behaviors. Instagram posts are collected using drug related terms by analyzing the hashtags supplied with each post. A large and dynamic dictionary of frequent illicit drug related slangs is used to find these posts. These posts are studied to find common drug consumption behaviors with regard to time of day and week. Furthermore, by studying the accounts followed by the users of drug related posts, we hope to discover common interests shared by drug users.

## 1 Introduction

Traditionally, online surveys, hard-copy surveys and telephone surveys are used to conduct studies related to risky behavior. However, surveys are not only time-consuming but may also suffer from sampling noise and small scale [3]. Therefore, we need to find a new and effective method for studying risky behaviors. As of 2015, there are approximately 2.2 billion social media users [4]. With the growing availability of the Internet, this number is only going to grow further. Social media is steadily becoming a more mainstream part of our lives and contains large amounts of invaluable data that can be analyzed. The NSDUH, a national wide survey that aims to collect data on people's drug consumption habits, only had 67,901 respondents in 2014 [11], which is insignificant compared to the number of social media users in the country. As social media becomes more popular, people start posting on every facet of life on them. This led us to choose social media as our choice of data source for this study. In particular, we chose Instagram over other social media platforms due to its popularity among young people [13]. Drug abuse is a major problem for society, and the menace has been steadily growing. According to the National Institute of Drug Abuse, "substance abuse costs our society more than 484 billion dollars a year," which is about thrice of what we spend on cancer. In 2000, there were about 460,000 deaths caused by illicit drug abuse [5]. Therefore, drug abuse presents a real threat to society and we need more effective methods to combat such a threat [12]. In this paper, we propose a novel approach to studying drug using patterns and networks by mining social media.

The main contributions of this study are as follows:
1. Establishing Instagram as a reliable data source to study patterns and trends related to certain illicit drugs;
2. Employing item set matching algorithms to understand new linguistic trends related to drug culture;
3. Compiling drug consumption behavior patterns with respect to time; and
4. Mining drug-user accounts to discover common interests shared by drug users;

## 1.1 Related Work

A number of studies on mining social media to discover behavior patterns have been conducted by researchers, examples being: predicting depression via social media [1], identifying disease outbreak by analyzing tweets [6], catching the criminals by using Facebook [7] and analyzing alcohol-related promotions on Twitter [8]. Additionally, Buntain and Golbeck [2] mined tweets from Twitter to discover drug use patterns. While the findings in [2] also pertain to time and location related information, our study is directed at discovering *more fine-grained, more repeatable behavior patterns*. Also, we have developed a technique that could allow us to keep updating our drug-related hashtags database, because drug related lingo is ever changing.

## 2 Methods

Instagram was the choice of social media for this paper. Instagram provides a RESTful API and we utilized this along with PHP's Guzzle library to make requests to the Instagram servers to identify and retrieve details of posts we deemed drug related. To identify posts that were drug related, we searched the hashtags attached with the posts, to see if the post contained any sensitive hashtags that we deemed drug associated. For this, we compiled a library of drug related hashtags, and these hashtags were identified using a dataset provided by the New York State Attorney General's office (NYSAGO). The dataset provided by the NYSAGO has 1,000 posts that were classified under human supervision as drug related and contains all the hashtags, image, comments of the posts. We used the 100 most frequent hashtags found in posts classified as drug related. We then used the recent endpoint to search for posts containing these sensitive hashtags and compiled a database of these posts. To increase the accuracy, we only fetch the posts with two or more matching hashtags, for 58,512 posts in total. Moreover, from those posts, we also fetched 100 drug users' accounts information (followers and followed-by) to study their common interests.

## 2.1 Drug Use Pattern Analysis

There are three major kinds of illicit drugs that have been studied in this paper: weed (cannabis), cough syrup (purple drank), and prescription pills of various kinds like Vicodin that are frequent abused by people. By separating the hashtags into those three categories under human supervision, we were able to identify posts about different drugs. In order to increase the accuracy, only the posts with more than 80% of hashtags that are about a certain drug were considered as related to this drug.

## 2.2 Linguistic Analysis

Since Instagram has been banning sensitive hashtags for drug use (such as '#dirtysprite'), drug-related hashtags are ever changing. In order to keep our hashtag set up-to-date, we are using the Apriori algorithm [9] to mine frequently used hashtag sets. First, the initial hashtag set was used to fetch more drug related posts (according to the Instagram API policy, we can fetch more than 50,000 posts per hour and approximately 100 of them are considered as drug related) and since the information of post is stored in JSON format, we can easily get a list of hashtags used in this post by using Python Dictionary. Basically, we treated each post as a 'transaction' of Apriori and each hashtag as an 'item' in the 'transaction'. Then, we stored the data in CSV files with one 'transaction' each row and hashtags separated by comma. Finally, we applied our Apriori Algorithm to those transactions and generated frequent item sets to update our hashtags dataset.

## 3 Results

### 3.1 Drug Popularity for Different Drugs and Time pattern

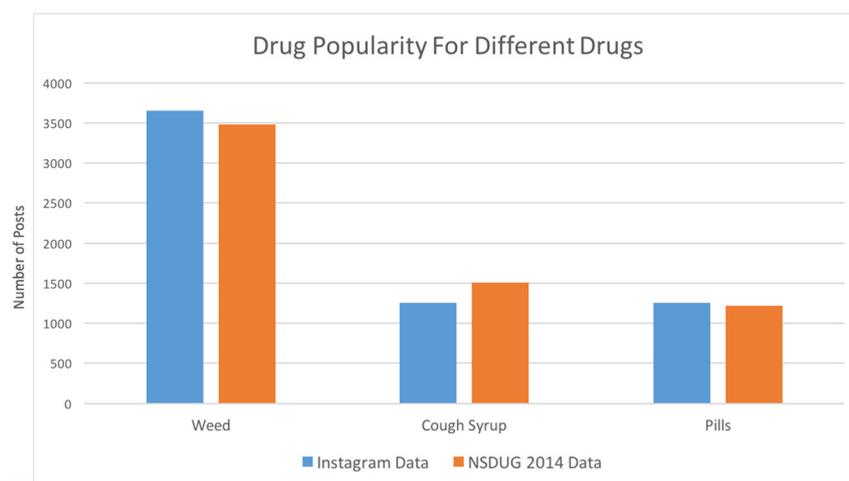

**Figure 1.** The comparison between the results by mining Instagram and the results from NSDUG 2014 survey data (rescaled to the same proportion).

The result for drug popularity is in the Figure 1. According to National Survey on Drug Use and Health 2014 [10], there are about 72% samples about consuming cannabis in 2014, 14% consuming pills and 13% consuming cough syrup. We plotted those data as orange bars in Figure 1. Therefore, as we can see from Figure 1, this survey basically matches the pattern we found: weed is the most popular, the consumption of pills and cough syrup are much less.

We used timestamp of each post to study the drug consumption patterns. Instagram records the GMT time for each post regardless of the time zone it was created in, which does introduce some errors while studying consumption patterns. However, since we only searched Instagram for English hashtags that we popular terms to refer to drugs in the United states and Europe, we can assume that majority of our posts are from places that are relative close to Greenwich meridian, hence this does not introduce any significant error in our analysis. After using the timestamps to study frequency of consumption, we found a number of interesting patterns related to drug use, and these are as follows (Figure 2 and Figure 3).

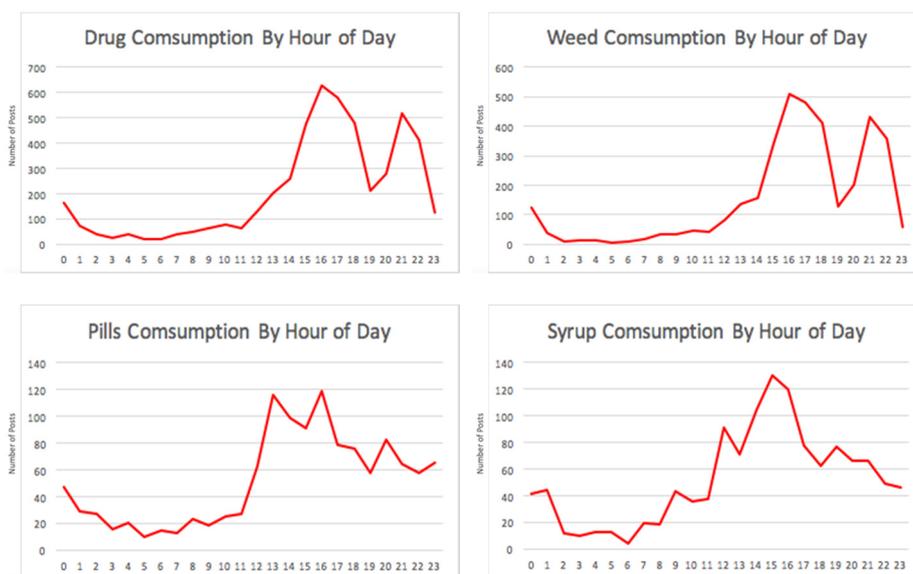

**Figure 2.** The time patterns of different illicit drug uses (by day).

Regarding hour of day, we see two peaks of consumption for weed, one at 16:00 hours and another at 21:00 hours. This seems legitimate because there exists a '420' culture in the weed community, which involves the act of consuming weed at 4:20 PM in the day. Also, since we have GMT timestamps, one peak corresponds to the 4:20 movement in GMT time zones, and the next peak corresponds to the 4:20 movement in the North American continent. Moreover, the consumption of drugs seems to increase towards the end of the day. Regarding day of week, people are more likely to consume drug on Thursdays than any other days (part of the reason is that people tend to have parties on Thursdays). Moreover, the consumption on Fridays is surprisingly higher than that on weekend days. Conventional wisdom seems to validate these findings, which further establishes Instagram as a valid way to study drug consumption patterns.

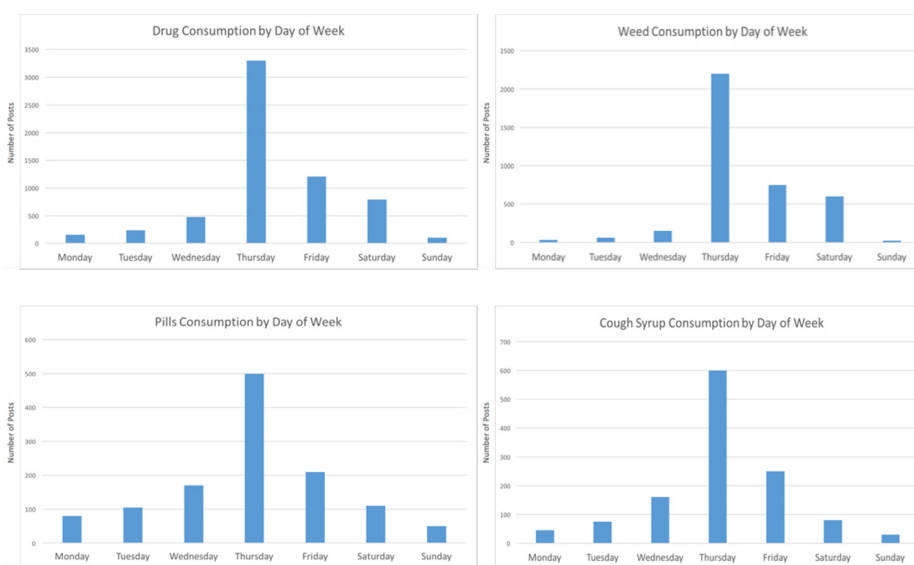

**Figure 3.** The time patterns of different illicit drug uses (by week).

### 3.2   Linguistic Analysis and Dynamic Hashtags Set

Those frequent hashtag sets are sorted based on their support. We then examine the top frequent drug-related hashtags and if the hashtag set, say, has a support over 20% (which is a relatively high support threshold in our case) and also is not in our hashtag database, then we add the hashtags in the set to our database.

During the mining process of frequent hashtag sets, we found some very interesting drug-related words or lingo. Some words that are drug-related were not used to describe drug use before, for examples:

- 'highsociety' has a support of more than 10% but was used to describe 'luxury life' instead of 'getting high'
- with support 5%, 'faded' that used to tell a feeling of 'unconsciousness' now means drug consumption
- 'poup' was meaning 'mix up' but now means 'make some cough syrup'

A common challenge with flagging sensitive posts is that language is constantly changing, with new words being added to the drug vocabulary and words having different meanings in drug related context, as shown above. Hence, dictionaries compiled quickly become outdated because of ever changing language. Using our approach, we have found a way to keep the drug related terms dictionary up to date, by adding the frequently occurring terms mined using the Apriori algorithm. However, this does involve human supervision, because some tags that appear with drug related tags are not drug related themselves, #goodtime for example. Hence, human supervision is required to select which hashtags to add to the dictionary.

### 3.3 Common Interests of Drug Related Accounts

After getting all the accounts followed by our 100 sample drug related accounts, we wanted to see if they had common interests, i.e., whether they all followed some common accounts. In order to do this, we ran the Apriori algorithm on the account names in the network and manually checked the top 10 accounts with the highest support. The results are as follows:

- Drug related accounts tend to follow glass-makers (people making devices for consuming drugs). Example: 'elboglass'(support 0.129),'saltglass'(support 0.107)
- Weed-related pages. Example: 'weedhumor'(support 0.102),'hightimesmagazine' (support 0.156)
- Celebrities. Example: 'christucker' (Chris Tucker), 'therock' (WWE wrestler and actor The Rock), 'heytommychong' (comedian) and 'cheechandchong' (comedian)

Some quite interesting association rules include:

- ('sdryno','coylecondenser')==>('oilbrothers','elksthatrun') with confidence 0.6. 'elksthatrun' is a music band that is followed by a lot of drug users, which may mean that drug users tend to like a certain kind of music.
- ('cheechandchong')==>('heytommychong','hightimesmagazine') with confidence 1, which implies that people who like comedian 'cheechandchong' also like 'hightimesmagazine'.

Drug related accounts seem to share common interests and have a similar network structure.

## 4    Conclusion and Future Work

In this paper, we proposed an effective way to update sensitive-term dictionaries for illicit drugs, which is important because of the ever changing lingo used to refer to various drugs. We mined common interests of drug related accounts and discovered that drug related accounts tend to follow certain kinds of celebrities, certain kinds of glass makers and that their network characteristics differ from those of non-drug related accounts. We were also able to mine association rules for drug related accounts, which said that if a drug related user follows account X, then he must also follow account Y. We hope this sheds light on the networking behaviors of drug related accounts.

We can further improve our analysis once we build classifiers that can differentiate drug dealers and users, as well as glass makers and legitimate sellers, and then use network analysis to study the social interactions among drug related users compared to non-drug related users.

**Acknowledgments.** This work was supported in part by New York State through the Goergen Institute for Data Science at the University of Rochester. We thank Lacy Kelly and Meredith McCarron of the NYSAGO for their valuable input to this study.